\documentclass[a4paper,12pt]{article}

\usepackage{typearea}
\typearea{15}

\usepackage{mymacros}

\usepackage{mathtools}
\usepackage{slashed}

\begin{document}

	\begin{center}
		{\LARGE Anisotropic Einstein universes with a global magetic field and SqK-spinors}\\
		\bigskip\bigskip
		{\large
		Fumihiro Ueno\footnote{ueno-fumihiro-dx@alumni.osaka-u.ac.jp}\\
			Satsuki Matsuno\footnote{smatsuno43@gmail.com}
		}\\
		\bigskip
		{\it Department of Physics, Graduate School of Science,
			Osaka Metropolitan University\\
			3-3-138 Sugimoto, Sumiyoshi, Osaka 558-8585, Japan}
	\end{center}
	
	
	\begin{abstract}
		We consider an Einstein-Dirac-Maxwell system with two charged massless spinors coupled with an electromagnetic field, and construct a family of exact solutions to the system.
  The solution spacetime is an anisotropic generalization of the static Einstein universe which has a global cosmic magnetic field generated by the current of the spinors.
  The spacetime is regarded as a toy model which describes global cosmic magnetic phenomena in the universe.
	The spinors are induced from Sasakian quasi-Killing spinors, and the total Dirac current flows along fibers of the Hopf-fibration.
 The magnetic field is a contact magnetic field.
	\end{abstract}
	
	\tableofcontents
	
	\section{Introduction}
	In the real universe, there is a somewhat strong global magnetic field \cite{Kulsrud 2008}, and the field may be induced by an electric current.
	Such a system is composed of Einstein gravity, an electromagnetic field, and matter generating the current.
	The system is often very complicated, and it is difficult to find exact solutions without simplification.
	
	So, it is interesting to constructing exact solutions to an Einstein-Maxwell system with an electric current.
	Many researchers have constructed exact solutions to Einstein-Maxwell systems with charged perfect fluids \cite{exact solutions}.	
	Ishihara and Matsuno (one of the present authors) also constructed a family of exact solutions to an Einstein-Maxwell system with a charged dust fluid on a static spacetime \cite{Ishihara Matsuno 2020,Ishihara Matsuno 2022.2}.
 
 Since fluid is a composite object, it is expected that there exists solutions composed of more fundamental matter, for example spinors, which describe similar situations.
 Then we consider an Einstein-Dirac-Maxwell (E-D-M) system.

	In this paper, we consider an E-D-M system consisting of two charged massless spinors $ \psi^\pm$ coupled with a gauge field $ A$ and the electromagnetic field $F=dA$ and Einstein graivty.
	The Lagrangian is given by
 \begin{align}
     \mathcal{L}= R-\frac{1}{4}F_{ab}F^{ab} - \sum_{i=+,-} \Re \lb \overline{\psi^i}\slashed{D}^i\psi^i \rb  -2 \Lambda. \label{eq:Lagrangian0}
 \end{align}

	Our spacetime ansatz is a static spacetime which is direct product spacetime of time and a Berger sphere.
 The spacetime is regarded as an anisotropic generalization of static Einstein universe.
 We construct a family of solutions which describe a situation where the Dirac-current of the spinors flows uniformly along fibers of Hopf-fibration of the Berger sphere and the current generates a magnetic field.
 The situation is slightly similar to of \cite{Dzhunushaliev}.
 
	The magnetic field is a contact magnetic field whose gauge field is proportional to a contact form, which was introduced in \cite{Cabrerizo 2009}.
	The motion of charged particles in contact magnetic fields and other geometrical properties have been studied in \cite{Cabrerizo 2009,Romaniuc 2015,Inoguchi Munteanu 2022} (and references therein).
	However, few studies have investigated the relationship between a contact magnetic field and Einstein gravity.
 
 The two massless spinors are induced from Sasakian quasi-Killing (SqK) spinors.
 	An SqK-spinor is a generalization of a Killing spinor in Sasakian geometry, which was introduced by Friedrich and Kim \cite{Kim Friedrich 2000}.
	Killing spinors play an important role in supergravity and have been actively studied \cite{Friedrich Thomas 2003,Semmelmann 1998,Fujii 1886,Felipe 2003} (and references therein).
	On the other hand, Sasakian quasi-Killing spinors seem to have few applications in mathematical physics.


	
	The organization of this paper is as follows.
 At the \hyperref[subsc:relusts]{end} of this section, we summarize our results.
	In Section \ref{sc:The Einstein-Dirac-Maxwell system} we describe our system in detail.
	Section \ref{sc:Solutions to the Einstein-Dirac-Maxwell system} is devoted to solving our system.
 In Appendix \ref{app:sasaki}, we briefly summarize Sasakian geometry for reader's convenient.

 \subsection*{Summary of our results}\label{subsc:relusts}
 We construct a familiy of exact solutions of the system \eqref{eq:Lagrangian0} as follows.
 Let $(M,g)$ be a spacetime whose metric is given by
 \begin{align}
		g&=-dt^2+\frac{\alpha}{4}r^2(d\theta^2+\sin^2\theta d\varphi^2)+\frac{\alpha^2}{4}r^2(d\phi+\cos\theta d\varphi)^2,\ \alpha\in\mathbb{R}_{\geq0},
	\end{align}
 and we introduce an orthonormal frame
 	\begin{align}
 X_0&=\partial_t,\ 
		X_1=\frac{2}{r\alpha}\partial_\phi,\\
		X_2&=\frac{2}{r\sqrt{\alpha}}\left(-\sin\phi\partial_\theta+\frac{\cos\phi}{\sin\theta}\partial_\varphi-\cot\theta\cos\phi\partial_\phi\right),\\
		X_3&=\frac{2}{r\sqrt{\alpha}}\left(\cos\phi\partial_\theta+\frac{\sin\phi}{\sin\theta}\partial_\varphi-\cot\theta\sin\phi\partial_\phi\right).
	\end{align}
 
 Let $A=\frac{2B}{\alpha}(d\phi+\cos\theta d\varphi)$ be a gauge field, and suppose that two charged massless spinors are given by
\begin{align}
    \psi^\pm=e^{\sqrt{-1}Et}C^\pm\begin{pmatrix}
        1 \\ \pm1 \\ 0 \\ 0
    \end{pmatrix},\ C^\pm\in\mathbb{C},
\end{align}
with respect to the frame $\{X_0,X_1,X_2,X_3\}$, and assume that they have charges $\pm q$ respectively.
The quantities $\Lambda,E,C^\pm,B,q$ are given by
 \begin{align}
\Lambda&= \frac{H^{2} + 6 H + 11}{3 r^{2} \left(H + 3\right)},\ E= \frac{H^{2} + 6 H + 11}{12 r},\ |C^\pm|^2= \frac{2}{r \left(H + 3\right)},\\
B^2 &= \displaystyle \frac{r^{2} \left(H - 1\right) \left(H + 4\right)}{6 \left(H + 3\right)},\ 
q^2 = \displaystyle \frac{\left(H - 1\right) \left(H + 3\right) \left(H + 4\right)}{24 r^{2}},
 \end{align}
  where $H:=\frac{4}{\alpha}-3$ is the sectional curvature of the plane-section spaned by $X_2$ and $X_3$.
  
 On these solutions, the total Dirac current flows uniformly along the orbit of $X_1$ which is tangent to a fiber of the Hopf fibration of Berger sphere $S^3$.
 And then the current generates a global cosmic magnetic field $F=dA$.

	\section{Building the universe model with charged spinors}\label{sc:The Einstein-Dirac-Maxwell system}
	We describe our E-D-M system on a four-dimensional spacetime $(M,g)$.
 We consider two charged massless spinors $ \psi^\pm $ coupled with a gauge field $ A_a $ and assume that each $ \psi^\pm $ has a charge $\pm q$ respectively.
	The Lagrangian is given by
	\begin{equation}\label{eq:Lagrangian density of E-D-M}
		\mathcal{L}= R-\frac{1}{4}F_{ab}F^{ab} -  \Re \lb \overline{\psi^+}\slashed{D}^+\psi^+ \rb  - \Re \lb \overline{\psi^-} \slashed{D}^-\psi^- \rb -2 \Lambda,
	\end{equation} 
	where $R$ is the scalar curvature, $F=dA$ is the electromagnetic field, $\overline{\psi}=\psi^\dagger\gamma_0$ is the Dirac conjugate, $ \slashed{D}^{\pm}=\sqrt{-1}\gamma^\mu(\nabla_\mu\pm \sqrt{-1}qA_\mu) $ are the gauged Dirac operators, and $ \Lambda $ is a cosmological constant.
	Then the Euler-Lagrange equations are given by
	\begin{align}
		&R_{ab} -\frac{1}{2}Rg_{ab} + \Lambda g_{ab} = T^{\rm em}_{ab}+T^{\rm spin}_{ab}, \label{eq:Einstein}\\
		&\slashed{D}^\pm\psi^\pm = 0, \label{eq:dirac}\\
		&\nabla^aF_{ab}  = q(\overline{\psi^+} \gamma^\mu \psi^+-\overline{\psi^-} \gamma^\mu \psi^-)(X_\mu)_b\label{eq:maxwell},\\
  &T^{em}_{ab}=F_a^{~c}F_{bc}-\frac{1}{4}F_{cd}F^{cd}g_{ab},\\
  &T^{\rm spin}_{ab}=T^{\psi^+}_{ab}+T^{\psi^-}_{ab},\\
&T^{\psi^\pm}_{ab}=\frac{1}{4}\Re(\overline{ \psi^\pm}\sqrt{-1}(\gamma_a\nabla_b^{\pm q}+\gamma_b\nabla_a^{\pm q})\psi^\pm).  
	\end{align}
	where $ \{X_\mu\}_{\mu=0}^3 $ is a global orthonormal frame and $\gamma_a=\gamma^\mu\otimes(X_\mu)_a$, and $\Re$ denotes the real part.
	
Let $ (N, h) $ be a three-dimensional oriented Riemannian manifold, and our spacetime $ (M, g) $ be a four-dimensional static spacetime defined by $ M = \bR \times N $ and $ g = -dt^2 + r^2h ,$ where $ t $ is a coordinate of $\bR$ and $ r $ is a positive constant.
We assume that the two charged spinors $\psi^\pm$ are given by
$$
\psi^\pm(t,x)=e^{\sqrt{-1}Et}\begin{pmatrix}
    \phi^\pm(x) \\ 0
\end{pmatrix},\ (t,x)\in\mathbb{R}\times N,
$$
for the chiral representation, where $\phi^\pm$ are spinors on $N$, and $E$ is a real constant.

Furthermore, our ansatz of the 3-space $(N,h)$ is a three-dimensional simply connected Sasakian manifold with constant scalar cuvature, and let $\xi^a$ and $\eta_a$ be the Reeb vector field and the contact 1-form respectively.
We briefly summarize Sasakian geometry in Appendix \ref{app:sasaki} for reader's convenient.
Then we put a gauge potentioal 
\begin{align}
 A_a=B\eta_a,\ B\in\mathbb{R},
\end{align}
that is a contact electromagnetic field \cite{Cabrerizo 2009}.
This is regarded as a global cosmic magnetic field.
It shall becomes clear that $(N,h)$ is a Berger sphere, but for the sake of generality we start our discussion in the above settings, that is, $(N,h)$ is a Sasakian 3-manifold.

	\section{Solving the model}\label{sc:Solutions to the Einstein-Dirac-Maxwell system}
	In this section, we solve our E-D-M system consisting of the Einstein equation \eqref{eq:Einstein}, the Dirac equation \eqref{eq:dirac} and the Maxwell equation \eqref{eq:maxwell}.
 Hereafter we take Clifford algebra representation as follows.
 The Pauli matrices $ \sigma_i$ are given by
	\begin{equation}
		\sigma_1
		=\begin{pmatrix}
			0 & 1 \\
			1& 0
		\end{pmatrix},
		\sigma_2
		=\begin{pmatrix}
			0 & -\sqrt{-1} \\
			\sqrt{-1}& 0
		\end{pmatrix},
		\sigma_3
		=\begin{pmatrix}
			1 & 0 \\
			0& -1
		\end{pmatrix},
	\end{equation}
	and the gamma matrices of the chiral representation are written as
	\begin{equation}\label{eq:Gamma matrics}
 \gamma_0 = 
		\begin{pmatrix}
			0 & I_2 \\  I_2 & 0
		\end{pmatrix},\ 
		\gamma_i = 
		\begin{pmatrix}
			0 & \sigma_i \\  -\sigma_i & 0
		\end{pmatrix} (i=1,2,3).
	\end{equation}
 We use a matrix representation of Clifford algebra $Cl_{1,3}$ written as
 $$
 X_\mu\mapsto \gamma_\mu, \ (\mu=0,1,2,3),
 $$
 for an orthonormal frame $\{X_\mu\}$ on $M$, and that of $ Cl_{0,3}  $ written as
	\begin{equation}\label{eq:Clifford-three-rep}
		e_i  \mapsto s_i\coloneqq\sqrt{-1}\sigma_i \ (i=1,2,3),
	\end{equation}
	for an orthonormal frame $ \{e_i\} $ on $ N $.
 We often write $\gamma^\mu=\eta^{\mu\nu}\gamma_\nu$.

 \subsection{Curvatures}
 Simply connected three-dimensional Sasakian manifolds $(N,h)$ with constant scalar curvature are homogeneous and there are three kinds.
 Let $H$ be the sectional curvature of the plane-section orthongonal to the Reeb vector $\xi$.
 The Sasakian space $(N,h)$ is in  Bianchi type IX if $ H>-3$, type II if $ H=-3$, or type VIII if $ H<-3$.
	They are diffeomorphic to $ {\rm S}^3,{\rm Nil}_3,\widetilde{{\rm SL}(2,\mathbb{R})}$ respectively.
 There is a useful global orthonormal frame on $(N,h)$, that is, 
	there exists a global orthonormal frame $ \{e_1=\xi,e_2,e_3\}$ which satisfies 
	\begin{align}
		[e_1,e_2]=-\frac{H+3}{2}e_3,\ [e_1,e_3]=\frac{H+3}{2}e_2,\ [e_2,e_3]=-2e_1.\label{eq:sasakian frame}
	\end{align}
	We call this frame a Sasakian frame (Appendix \ref{app:sasaki}).
	The Ricci tensor of $ (N,h)$ is given by 
	\begin{align}
		{\rm Ric}_N={\rm diag}(2,H+1,H+1),
	\end{align}
	with respect to a Sasakian frame $ \{e_1,e_2,e_3\}$.

 We also define a global orthonormal frame $ \{X_0,X_1,X_2,X_3\}$ on $ (M,g)$, that is
\begin{align}
    X_0=\partial_t,\ X_i=\frac{1}{r}e_i,\ (i=1,2,3).
\end{align}
We also call this frame a Sasakian frame.
Moreover the Ricci tensor of $ (M,g)$ is given by 
	\begin{align}
		{\rm Ric}_M={\rm diag}\left(0,\frac{2}{r},\frac{H+1}{r},\frac{H+1}{r}\right),
	\end{align}
	with respect to a Sasakian frame $ \{X_0,X_1,X_2,X_3\}$.

\subsection{Ansatz of spinors}
We use a Sasakian quasi-Killing spinor to solve our E-D-M system.
On a Sasakian space $(N,h)$, a Sasakian quasi-Killing (SqK) spinor of type $(a,b)$ is a spinor $\phi$ which satisfies
		\begin{equation}\label{eq:Sasakian-quasi-Killing}
			\nabla^N_{e_i} \phi = a s_i \phi + b\delta_{1i}s_i\phi,
		\end{equation}
  where $ a,b$ are real numbers, $\{e_1=\xi,e_2,e_3\}$ is a orthonormal frame, and $\nabla^N$ is the spin connection of $(N,h)$.
An SqK-spinor is defined by Friedrich and Kim\cite{Kim Friedrich 2000} and they show that an SqK-spinor of type $\left(\frac{1}{2},\frac{H-1}{4}\right)$ exists on an arbitrary simply connected three-dimensional Sasakian manifold with constant scalar curvature.
The SqK-spinor is a Killing spinor if $H=1$ case, that is, $(N,h)$ is the round $S^3$.
Some physical properties and explicit formulae of SqK-spinors in three-dimension are discussed in \cite{MatsunoUeno}.

Let $\phi^\pm$ be SqK-spinors of type $\left(\frac{1}{2},\frac{H-1}{4}\right)$ such that
\begin{align}
\xi\cdot\phi^\pm=\pm\sqrt{-1}\phi^\pm. \label{eq:As1}
\end{align}
For an SqK-spinor $\phi$, the condition \eqref{eq:As1} is equivalent to that the current $\sum_i(\phi^\dagger s_i\phi)e_i$ is a Killing vector field \cite{MatsunoUeno} proportional to $\xi$.
We assume that two charged massless spinors $\psi^\pm$ are of the form
\begin{align}
    \psi^\pm=e^{\sqrt{-1}Et}\begin{pmatrix}
    \phi^\pm \\ 0
    \end{pmatrix}.
\end{align}
Since \eqref{eq:As1}, we have $\gamma_1\psi^\pm=\pm\gamma^0\psi^\pm$.

With respect to a Sasakian frame $\{e_i\}$, the spin connection $\nabla^N$ is given by
\begin{align}
    \nabla^N=d+\frac{1}{4}\sum_{i,j,k}h(\nabla_{e_i}e_j,e_k)\omega^i\otimes s_js_k
			=d+\frac{1}{2}\sum_k \omega^k\otimes s_k+\frac{H-1}{4}\eta\otimes s_1,
\end{align}
it follows that an SqK-spinor $\phi$ of type $\left(\frac{1}{2},\frac{H-1}{4}\right)$ is obviously given by
\begin{align}
    \phi=\begin{pmatrix}C_1 \\ C_2\end{pmatrix},\ C_1,C_2\in\mathbb{C},
\end{align}
therefore, since $\xi\phi^\pm=\pm\sqrt{-1}\phi^\pm$, we have the component representation
\begin{align}
    \phi^\pm=C^\pm\begin{pmatrix}1 \\ \pm1\end{pmatrix},\ C^\pm\in\mathbb{C}.
\end{align}


\subsection{The Dirac-Maxwell equation}
	Here we explicitly write down the Maxwell equation \eqref{eq:maxwell}.
 We denote the co-frame of $ \{e_1,e_2,e_3\}$ by $ \{\omega^1=\eta,\omega^2,\omega^3\}$.
	Since $ A=B\eta $ and $F=dA,$ we have
	\begin{align}
		\ast d\ast F&=\ast d\ast(2B\omega^2\wedge\omega^3)
		=\ast d\left(\frac{2B}{r}\omega^0\wedge \omega^1\right)\nonumber\\
		&=\ast\left(-\frac{4B}{r}\omega^0\wedge\omega^2\wedge\omega^3\right)
		=\frac{4B}{r^2}\omega^1,
	\end{align}
 where we used $d\omega^1=2\omega^2\wedge\omega^3$ which is derived from \eqref{eq:sasakian frame}.
	It follows that the Maxwell equation \eqref{eq:maxwell} is given by
	\begin{align}
		\frac{4B}{r^3}X_1&=q(\overline{\psi^+} \gamma^\mu \psi^+-\overline{\psi^-} \gamma^\mu \psi^-)X_\mu\\
  &=q(||\phi^+||^2-||\phi^-||^2)X_0-q(||\phi^+||^2+||\phi^-||^2)X_1,
	\end{align}
	where $||\phi||^2:=\phi^\dagger\phi$ for a spinor $\phi$ on $N$.

 Then we can put $p^2:=||\phi^+||^2=||\phi^-||^2$, and we have
 \begin{align}
     p^2=-\frac{2B}{qr^3} \label{eq:maxwell 2}.
 \end{align}
Since $\sum_is_i\nabla^N_{e_i}\phi^\pm=-(3a+b)\phi^\pm=-\frac{H+5}{4}\phi^\pm$ holds, then the Dirac equations for $\psi^\pm$ are given by
\begin{align}
    \slashed{D}^\pm\psi^\pm&=\sqrt{-1}\gamma^\mu(\nabla^M_{X_\mu}\pm \sqrt{-1}qB\eta(X_\mu))\psi^\pm\\
    &=-E\gamma^0\psi^\pm-\frac{1}{r}\gamma^0\begin{pmatrix}s_i\nabla^N_{e_i}\phi^\pm\\ 0\end{pmatrix}-\frac{qB}{r}\gamma^0\psi^\pm\\
    &=\gamma^0\left(-E+\frac{H+5-4qB}{4r}\right)\psi^\pm,
\end{align}
then we have
\begin{align}
    -E+\frac{H+5-4qB}{4r}=0. \label{eq:dirac2}
\end{align}

\subsection{Energy tensors}
By simple calculation, we have the energy-momentum tensor of $F$ as follows
	\begin{align}
		T^{\rm em}=\frac{2B^2}{r^4}{\rm diag}(1,-1,1,1),\label{eq:energytensor of electromagnetic 2}
	\end{align}
	with respect to a Sasakian frame $\{X_0,X_1,X_2,X_3\}$.
 
Next we calculate the energy-momentum tensor of $\psi^\pm$.
We have 
\begin{align}
    		\nabla_{X_0}^{\pm q}\psi^\pm &=\sqrt{-1}E\phi^\pm e^{\sqrt{-1}Et}\oplus 0,\\
   \nabla_{X_i}^{\pm q}\psi^\pm &=\sqrt{-1}\lc \frac{1}{2r}\sigma_i + \lb \frac{H-1}{4r}\sigma_1 \pm \frac{qB}{r} \rb \delta_{1i} \rc \phi^\pm e^{\sqrt{-1}Et}\oplus 0,
\end{align}
			
then it follows that
		\begin{align}
			\overline{\psi^\pm}\sqrt{-1}X_0 \cdot \nabla^{\pm q}_{X_0} \psi^\pm & = E||\phi^\pm||^2\\
			\overline{\psi^\pm}\sqrt{-1}X_0 \cdot \nabla^{\pm q}_{X_i} \psi & =(\phi^\pm)^\dagger \lc \frac{1}{2r}\sigma_i + \lb \frac{H-1}{4r}\sigma_1 \pm \frac{qB}{r} \rb \delta_{1 i} \rc \phi^\pm,\\ 
			\overline{\psi^\pm}\sqrt{-1}X_i \cdot \nabla^{\pm q}_{X_0} \psi & = E(\phi^\pm)^\dagger \sigma_i \phi^\pm, \\ \notag
			\overline{\psi^\pm}\sqrt{-1}X_i \cdot \nabla^{\pm q}_{X_j} \psi & = \frac{1}{2r}(\phi^\pm)^\dagger\sigma_i\sigma_j\phi^\pm + \lb \frac{H-1}{4r}(\phi^\pm)^\dagger \sigma_i\sigma_1\phi^\pm \pm \frac{qB}{r}(\phi^\pm)^\dagger \sigma_i \phi^\pm \rb\delta_{1j},\\
		\end{align}
which implies
\begin{align}
			T^{\psi^\pm}(X_0, X_0) & = \frac{Ep^2}{2} \\
			T^{\psi^\pm}(X_0, X_i) & = \frac{1}{4}(\phi^\pm)^\dagger\lc\left(\frac{1}{2r}+E \right)\sigma_i+\left(\frac{H-1}{4r}\sigma_1\pm\frac{qB}{r}\right)\delta_{1i}\rc\phi^\pm \\ 
			T^{\psi^\pm}(X_i, X_j) & =   \frac{p^2}{4r} \delta_{ij} + \frac{H-1}{8r}p^2\delta_{1i}\delta_{1j} \pm \frac{qB}{4r}\lb (\phi^\pm)^\dagger \sigma_i \phi^\pm\delta_{1j} + (\phi^\pm)^\dagger \sigma_j \phi^\pm\delta_{1i} \rb .
		\end{align}

Therefore $T^{\rm spin}=T^{\psi^+}+T^{\psi^-}$ is given by
\begin{align}
    T^{\rm spin}(X_0,X_0)&=Ep^2,\\
    T^{\rm spin}(X_1,X_1)&=\frac{H+1+4qB}{4r}p^2,\\
    T^{\rm spin}(X_2,X_2)&=T^{\rm spin}(X_3,X_3)=\frac{p^2}{2r},\\
    T^{\rm spin}({\rm else})&=0,
\end{align}
where we used $(\phi^\pm)^\dagger\sigma_2\phi^\pm=(\phi^\pm)^\dagger\sigma_3\phi^\pm=0$ since \eqref{eq:As1}.

\subsection{Solutions}
We have the Maxwell equation \eqref{eq:maxwell 2}, the Dirac equation \eqref{eq:dirac2} and the Einstein equation
\begin{align}
		\frac{2+H}{r^2}-\Lambda&=\frac{2B^2}{r^4}+T^{\rm spin}_{00},\label{eq:ein00}\\
		-\frac{H}{r^2}+\Lambda&=-\frac{2B^2}{r^4}+T^{\rm spin}_{11},\label{eq:ein11}\\
		-\frac{1}{r^2}+\Lambda&=\frac{2B^2}{r^4}+T^{\rm spin}_{22}\label{eq:ein22},
	\end{align}

 We regard $H$ and $r$ as parameters of our system.
 Solving the system, we obtain a family of solutions:
 \begin{align}
\Lambda&= \frac{H^{2} + 6 H + 11}{3 r^{2} \left(H + 3\right)},\ 
E= \frac{H^{2} + 6 H + 11}{12 r},\ 
p^2= \frac{4}{r \left(H + 3\right)},\label{eq:solution1}\\
B^2&= \frac{r^{2} \left(H - 1\right) \left(H + 4\right)}{6 \left(H + 3\right)},\ 
q^2= \frac{\left(H - 1\right) \left(H + 3\right) \left(H + 4\right)}{24 r^{2}}.\label{eq:solutions2}
 \end{align}

Since $p^2>0$, thus $H>-3$ holds, therefore $(N,h)$ is a Berger sphere.
Furthermore, since $B,q\in\mathbb{R}$, then we have $H>1$.
The cosmological constant $\Lambda$ and the energy eigenvalue $E$ is always positive.

Our solution spacetime $(M,g)$ is homeomorphic to $\mathbb{R}\times S^3$ and $(S^3,h)$ is a Berger sphere, so this is an anisotropic generalization of the static Einstein universe.
With respect to a suitable coordiante $(t,\theta,\phi,\varphi)$, the metric of solution spacetime is expressed as
	\begin{align}
		g=-dt^2+\frac{\alpha}{4}r^2(d\theta^2+\sin^2\theta d\varphi^2)+\frac{\alpha^2}{4}r^2(d\phi+\cos\theta d\varphi)^2,\ H=\frac{4}{\alpha}-3,
	\end{align}
 where $\alpha\in(0,1)$ is a parameter.
	A Sasakian frame is given by
	\begin{align}
 X_0&=\partial_t,\ 
		X_1=\frac{2}{r\alpha}\partial_\phi=\xi,\\
		X_2&=\frac{2}{r\sqrt{\alpha}}\left(-\sin\phi\partial_\theta+\frac{\cos\phi}{\sin\theta}\partial_\varphi-\cot\theta\cos\phi\partial_\phi\right),\\
		X_3&=\frac{2}{r\sqrt{\alpha}}\left(\cos\phi\partial_\theta+\frac{\sin\phi}{\sin\theta}\partial_\varphi-\cot\theta\sin\phi\partial_\phi\right).
	\end{align}

The gauge field is $A=\frac{2B}{\alpha}(d\phi+\cos\theta d\varphi)$ and the two charged massless spinors are given by
\begin{align}
    \psi^\pm=e^{\sqrt{-1}Et}C^\pm\begin{pmatrix}
        1 \\ \pm1 \\ 0 \\ 0
    \end{pmatrix},\ (C^\pm\in\mathbb{C},\ p^2=2|C^\pm|^2),
\end{align}
with respect to the Sasakian frame $\{X_0,X_1,X_2,X_3\}$.

	\section{Conclusions and discussions}\label{sc:Conclusions and discussions}
	We considered an E-D-M system consisting of Einstein gravity, an electromagnetic field and two charged massless spinors.
	We constructed a family of exact solutions to the system.
	The solution spacetimes are static direct product spacetimes of time $\mathbb{R}$ and Berger spheres $(S^3,h)$.
	The electromagnetic field is a contact magnetic field whose gauge field is proportional to the contact form of the Berger sphere.
	The two massless spinors have opposite charges and their total electric current flows along the Reeb orbits.
 The spinors are induced from Sasakian quasi-Killing spinors on the Berger sphere.
 
 Our solutions describe a situation in which the Dirac currents flow uniformly along fibers of Hopf fibration of Berger spheres, and the current generates a global cosmic magnetic field.
 From solutions \eqref{eq:solution1} and \eqref{eq:solutions2}, we can read the following.
 As the magnetic field strength becomes stronger, i.e., as B increases, the magnetic tension pulls in the direction of the fiber, causing the fiber to shrink and the Berger sphere to bend toward the basespace and become smaller as a whole.
 At the same time, the cosmological constant that supports the space so that it does not collapse also grows.

 In this paper we use charged massless spinors, but the authors could verify that charged massive spinors and a contact magnetic field also constitute a family of exact solutions to the E-D-M system.
 The calculations follow a similar process to that in this paper, but are considerably more complicated.

	\section*{Acknowledgement}
	We consulted H. Ishihara, K. Nakao and H. Yoshino for discussion on the weak energy condition.
	We would like to express our gratitude to them.

 This work was partly supported by MEXT Promotion of Distinctive Joint Research Center Program JPMXP0723833165.

	\appendix
\section{Sasakian manifolds}\label{app:sasaki}
We recall some basic facts in Sasakian geometry.
\begin{definition}\label{df:Sasaki mfd}
		A \textbf{contact metric manifold} is a quintuple $(M, \phi^a_{~b}, \xi^a, \eta_a, g_{ab})$ which satisfies the following conditions.
		\begin{align}
			(\phi^2)^a_{~b} &= -\delta^a_{~b} + \xi^a\eta_b,\\
			\eta_a\xi^a&=1, \\
			g_{cd}\phi^c_{~a}\phi^d_{~b} &= g_{ab} - \eta_a\eta_b,\\
			\phi_{ab} &= \nabla_a\eta_b-\nabla_b\eta_a
		\end{align}
		The quadruple $ (\phi^a_{~b},\xi^a, \eta_a, g_{ab}) $ is called a contact metric structure of $ M$.
		The one-form $ \eta_a $ and the vector field $ \xi^a $ are called the \textbf{contact form} and the \textbf{Reeb vector field} of the contact metric structure.
		A contact manifold $(M, \phi^a_{~b}, \xi^a, \eta_a, g_{ab})$ is called a \textbf{Sasakian manifold} if the Riemann curvature tensor $ R^a_{~dbc}$ satisfies $ R^a_{~dbc}\xi^d = \eta_c\delta^a_b-\eta_b\delta^a_c $.
	\end{definition}

 For an arbitrary vector field $X^a$ orthogonal to $\xi^a$, the vector field $\phi^a_{~b}X^b$ is orthogonal to $X^a$ and $\xi^a$.
 The sectional curvature of the plane-section spaned by $X^a$ and $\phi^a_{~b}X^b$ is called a \textbf{$\phi$-sectional curvature}.
 A complete simply connected Sasakian manifold which has constant $\phi-$sectional curvature is called a \textbf{Sasakian space form}.
In three-dimensions, the following useful condition is known \cite{KonYano}.

\begin{proposition}\label{pr:3 dim sasaki cond}
A three-dimensional Riemannian manifold $ (M,g) $ is a Sasakian manifold if and only if $ M $ admits a unit Killing vector field $ \xi^a $ and the value of the sectional curvature for a plane section including $\xi^a$ is equal to one.

	\end{proposition}

Since \pref{pr:3 dim sasaki cond}, the Ricci tensor of a three-dimensional Sasakian manifold is given by
$$
\textrm{Ric}=\textrm{diag}(2,H+1,H+1),
$$
with respect to an orthonormal frame $\{e_1=\xi,e_2,e_3\}$, where $H$ is the $\phi$-sectional curvature.
For three-dimensional Sasakian manifold, being a Sasakian space-form is equivalent to that the scalar curvature is constant.
The classification of three-dimensional Sasakian space-forms is also known \cite{BoyerGalicki,Blair}.

\begin{proposition}\label{pr:3 dim sasaki classification}
A three-dimensional Sasakian space-form whose $\phi$-sectional curvature $H$ satisfies $H>-3$ (resp $H=-3$ or $H<-3$) is the Lie group $S^3$ (resp ${\rm Nil}_3$ or $\widetilde{SL(2,\mathbb{R})}$) furnished with a suitable invariant metric.
	\end{proposition}

A Sasakian space-forms isomporphic to $S^3$ is called a \textbf{Berger sphere}.
There exists an orthonormal frame $\{e_1=\xi,e_2,e_3\}$ for each Sasakian space-form, which satisfies
	\begin{align}
		[e_1,e_2]=-\frac{H+3}{2}e_3,\ [e_1,e_3]=\frac{H+3}{2}e_2,\ [e_2,e_3]=-2e_1,\ H\in\mathbb{R}.
	\end{align}
one can easily verify the conditions in \pref{pr:3 dim sasaki cond} are satisfied, and the $\phi$-sectional curvature is $H$.
The frame gives a Sasakian frame \ref{eq:sasakian frame}.
The Lie algebra of $S^3,\ {\rm Nil}_3$ or $\widetilde{SL(2,\mathbb{R})}$ is of Bianchi type IX, II or VIII respectively.
The Lie algebra of Bianchi type IX, II or VIII is given by
\begin{align}
		[X_1,X_2]=-\alpha X_3,\ [X_1,X_3]=\alpha X_2,\ [X_2,X_3]=\beta X_1,
	\end{align}
	where $\alpha=-1,\beta=1$ for type IX, $\alpha=0,\beta=1$ for type II and $\alpha=\beta=-1$ for type VIII.
The Sasakian frame $\{e_1,e_2,e_3\}$ consists of a Lie algebra isomorphic to either of Bianchi type IX, II or VIII.

\end{document}